\documentclass[preprint,numbers,amsmath,superscriptaddress]{revtex4}
\usepackage{graphicx}
\usepackage{dcolumn}
\usepackage{bm}
\usepackage{epsfig}
\usepackage{multirow}
\usepackage{xcolor}

\begin{document}
\title{Transmission through correlated Cu$_n$CoCu$_n$ heterostructures} 
\author{L. Chioncel}
\address{Augsburg Center for Innovative Technologies, University of Augsburg,
D-86135 Augsburg, Germany}
\address{Theoretical Physics III, Center for Electronic Correlations and
Magnetism, Institute of Physics, University of Augsburg, 86135 Augsburg,
Germany}
\author{C. Morari}
\affiliation{National Institute for Research and Development of Isotopic and
Molecular Technologies, 65-103 Donath,  400293 Cluj Napoca, Romania}
\author{A. \"Ostlin}
\address{Department of Materials Science and Engineering, Applied Materials 
Physics, KTH Royal Institute of Technology, Stockholm 100 44, Sweden}
\author{W. H. Appelt}
\address{Augsburg Center for Innovative Technologies, University of Augsburg,
D-86135 Augsburg, Germany}
\address{Theoretical Physics III, Center for Electronic Correlations and
Magnetism, Institute of Physics, University of Augsburg, 86135 Augsburg,
Germany}
\author{ A. Droghetti}
\affiliation{Nano-Bio Spectroscopy Group and European Theoretical 
Spectroscopy Facility (ETSF), Universidad del Pais Vasco CFM
CSIC-UPV/EHU-MPC and DIPC, Av. Tolosa 72, 20018 San Sebastian
Spain}
\author{M. M. Radonji\'c}
\address{Theoretical Physics III, Center for Electronic Correlations and
Magnetism, Institute of Physics, University of Augsburg, 86135 Augsburg,
Germany}
\address{Scientific Computing Laboratory, Institute of Physics Belgrade,
University of Belgrade, Pregrevica 118, 11080 Belgrade, Serbia}
\author{I. Rungger}
\affiliation{School of Physics and CRANN, Trinity College, Dublin 2, Ireland}
\author{L. Vitos}
\address{Department of Materials Science and Engineering, Applied Materials 
Physics, KTH Royal Institute of Technology, Stockholm 100 44, Sweden}
\author{U. Eckern}
\address{Theoretical Physics II,
Institute of Physics, University of Augsburg, 86135 Augsburg, Germany}
\author{A. V. Postnikov}
\affiliation{LCP-A2MC, Institute Jean Barriol, University of Lorraine 
1, Bd Arago, 57078 Metz, France}

\begin{abstract}
The effects of local electronic interactions and finite temperatures upon the 
transmission across the Cu$_4$CoCu$_4$ metallic heterostructure are studied in 
a combined density functional and dynamical mean field theory. It is shown that, 
as the electronic correlations are taken into account via a local but dynamic 
self-energy, the total transmission at the Fermi level gets reduced (predominantly 
in the minority spin channel), whereby the spin polarization of the transmission 
increases. The latter is due to a more significant $d$-electrons contribution, as 
compared to the non-correlated case in which the transport is dominated by $s$ and
$p$ electrons. 
\end{abstract}

%\pacs{71.15.Ap;71.10.-w;73.21.Ac;75.50.Cc}

\maketitle

\section{Introduction}
The design of multi-layered heterostructures composed of alternating magnetic 
and non-magnetic metals offers large flexibility in tailoring spin-sensitive 
electron transport properties of devices in which the current flow is perpendicular 
to the planes. In the framework of ballistic transport, the spin-polarized 
conductance and the giant magnetoresistance effect (GMR) depends on the mismatch 
between the electronic bands of the concerned metals near the Fermi 
level~\cite{sc.ke.95, sc.ke.98}. In order to maximize the spin polarization 
of current and hence the GMR, heterostructures including half-metallic 
materials~\cite{gr.mu.83,zu.fa.04,ka.ir.08} seem to be the materials of choice.
In practice, however, the spin-polarization is never complete due to the 
presence of defects, and/or due to intrinsic limitations caused by 
spin-contamination and spin-orbit coupling~\cite{ka.ir.08}.
Owing to the technological relevance, considerable progress has been achieved 
in the computational description of multilayered heterostructures. In 
particular, the ballistic transport properties have been addressed by 
considering the Landauer-B\"uttiker 
formalism~\cite{land.57b,land.88,butt.86,butt.88}, where the conductance is 
determined by the electron transmission probability through the device region, 
which is placed between two semi-infinite electrodes. The transmission 
probability can be then computed with different electronic structure 
approaches, such as the 
tight-binding~\cite{sa.la.99,ma.um.97,ts.pe.01,kr.zh.02} or the 
first-principles density functional theory (DFT) 
ones~\cite{ho.ko.64,ko.sh.65,kohn.99}. Various implementations exist, 
based on transfer matrix~\cite{wo.is.02a,wo.is.02b}, 
layer-Korringa--Kohn--Rostoker (KKR)~\cite{st.ha.88,la.zh.99}, or 
non-equilibrium Green's function (NEGF)~\cite{ro.su.06} techniques.
 
In the context of first-principles calculations, it is known that, for 
systems with moderate to strong electron correlations, the electronic 
structure, calculated with the ``conventional'' DFT local density 
approximation (LDA) or its generalized gradient approximation (GGA) 
extension, is not accurate enough to account for the observed spectroscopic 
behavior. A more adequate description is provided within the dynamical mean 
field theory (DMFT)~\cite{me.vo.89,ge.ko.96,ko.vo.04} built on the LDA 
framework~\cite{an.po.97,li.ka.98,ko.sa.08}. Since many 
interesting magnetic materials fall into this category, the prediction of 
their electron transport properties, obtained by combining the 
Landauer-B\"uttiker formalism with DFT~\cite{ro.su.06,ro.su.05,ru.sa.08}, is 
expected to equally suffer from an insufficient treatement of correlation effects, 
which would be captured by adding DMFT. However, the full incorporation of 
correlation effects at the DMFT level into the transport calculations is not 
straightforward, not only because of technical reasons, such as large system 
size and lack of corresponding algorithms, but also because of conceptual 
difficulties in the development of many-body solvers in the out-of-equilibrium 
regime~\cite{ao.ts.14}. Attempts to close this gap include the use of combined 
techniques, in which equilibrium-DMFT calculations are performed in order to 
obtain the Landauer conductance of atomic contacts made of transition 
metals~\cite{ja.ha.09,ja.ko.10}.

In this work, we investigate the linear-response transport through a 
prototypical Cu-Co-Cu heterostructure by accounting for strong electron 
correlation effects in the electronic structure of the Co monolayer. The 
attention to this system is drawn by a significant density of states 
that develops in the Co layer, in the vicinity of the Fermi level, in 
one spin channel only (the minority-spin one), whereas the Cu layers 
contribute with states at higher binding energies. 
We use a  ``two-step'' approach, in which the Landauer transmission 
probability is calculated within the {\sc smeagol} NEGF based electron 
transport code~\cite{ro.su.06,ro.su.05,ru.sa.08} whereby the Hamiltonian 
is obtained from DFT~\cite{so.ar.02}. The many-body corrections to the 
Green's function are evaluated using DMFT in an exact muffin-tin orbitals 
(EMTO)-based package~\cite{an.sa.00,vi.sk.00,vito.01}, which uses a 
screened KKR approach~\cite{wein.90}. These corrections are then passed 
to {\sc smeagol} for the calculation of the transport properties.

The article is organized as follows. We start with a general description 
of the transport problem in the presence of electronic correlations
(Sec.~\ref{appr}). Then the computational details are outlined in 
Sec.~\ref{DMFT-calc}, and the geometry of the system considered in our 
simulations is descried in Sec.~\ref{geom}. Finally, Sec. \ref{results} 
presents the main results, and Sec. \ref{conclusion} summarizes and concludes. 
{The appendices deal with technical implementation and 
various tests, notably App.~\ref{basis_transf} explains porting the results
from EMTO into {\sc smeagol}, and discusses a model two-orbital system
of cubic symmetry.}

\section{Methods}
\label{method}
\subsection{Transport properties in the presence of electronic correlations}
\label{appr}
The electronic transport through a device can be addressed 
using the Kubo 
approach, where the central quantity is the conductivity, and the electrical 
current is the result of the linear response of the system to an applied 
electric field~\cite{mahan}. Alternatively, in the Landauer-B\"uttiker 
formulation~\cite{land.57b,land.88,butt.86,butt.88}, the current flow 
through a device is considered as a transmission process across a 
finite-sized scattering region placed between two semi-infinite leads,
connected at infinity to charge reservoirs.
The quantity of interest is the conductance, which, within linear response, is 
given by:
\begin{equation}
\mathcal{G}=\frac {e^2}{h}\frac{1}{\Omega_\mathrm{BZ}}\sum_{\sigma=\uparrow,\downarrow}\int_\mathrm{BZ}\!d\mathbf{k}_\parallel T_{\sigma}(\mathbf{k}_\parallel,E_F), \label{conductance}
\end{equation}
where $-e$ is the electron charge, $h$  the Planck constant, and $e^2/h$ 
half the quantum of conductance. $T_\sigma(\mathbf{k}_\parallel,E_F)$ 
is the spin-dependent transmission probability from one lead to the other 
for electrons at the Fermi energy and with the transverse wave-vector 
$\mathbf{k}_\parallel$ perpendicular to the current flow (here we assume that 
the two spin-components do not mix). The integral over 
$\mathbf{k}_\parallel$ goes over the Brillouin zone (BZ) perpendicular to 
the transport direction, and $\Omega_\mathrm{BZ}$ is the area of the BZ. In 
the case when the interaction between electrons involved in transport is 
completely neglected, the transmission for a given energy, $E$, of the 
incident electrons can be evaluated as~\cite{datta}
\begin{equation}
T_{\sigma}(\mathbf{k}_\parallel,E) = {\rm Tr} \left[
\mathbf{\Gamma}_{L}^\sigma (\mathbf{k}_\parallel,E) \mathbf{G}^{\sigma\dagger} (\mathbf{k}_\parallel,E) 
\mathbf{\Gamma}_{R}^\sigma (\mathbf{k}_\parallel,E) \mathbf{G}^{\sigma}(\mathbf{k}_\parallel,E)\right],
\label{landauer}
\end{equation}
where $\mathbf{G}^{\sigma}(\mathbf{k}_\parallel,E)$ is the retarded Green's 
function of the scattering region coupled to the leads,
\begin{equation}
\mathbf{G}^\sigma(\mathbf{k}_\parallel,E) = \left[
\epsilon^+\mathbf{S}(\mathbf{k}_\parallel) 
-\mathbf{H}^\sigma(\mathbf{k}_\parallel) 
-\mathbf{\Sigma}_{L}^\sigma(\mathbf{k}_\parallel,E) 
-\mathbf{\Sigma}_{R}^\sigma(\mathbf{k}_\parallel,E)
\right]^{-1}.
\label{Green_add}
\end{equation}
All terms presented are matrices $[\mathbf{G}^\sigma(\mathbf{k}_\parallel,E)]_{\mu\nu}$, 
labelled by the global indices  $\mu$, $\nu$ which run through the basis functions at all
atomic positions in the scattering 
region. $\mathbf{S}(\mathbf{k}_\parallel)$ represents the orbital overlap 
matrix, and the energy shift into the complex plane, 
$\epsilon^+ = \lim_{\delta \to 0^{+}}(E + i \delta)$, has been introduced 
to respect causality. $\mathbf{H}^{\sigma}(\mathbf{k}_\parallel)$ is the 
Hamiltonian of the scattering region for spin $\sigma$; the right and 
left self-energies 
$\mathbf{\Sigma}_{R}^\sigma(\mathbf{k}_\parallel,E)$ and 
$\mathbf{\Sigma}_{L}^\sigma(\mathbf{k}_\parallel,E)$ describe the energy-, 
momentum- and spin-dependent hybridization of the scattering region with 
the left and right leads, respectively \cite{ru.sa.08}. Therefore, 
$\mathbf{G}^\sigma(\mathbf{k}_\parallel,E)$  is formally the retarded Green 
function associated to the effective, non-hermitian Hamiltonian 
$\mathbf{H}^\sigma_{\rm eff}(\mathbf{k}_\parallel,E)=\mathbf{H}^\sigma(\mathbf{k}_\parallel) -\mathbf{\Sigma}_{L}^\mathbf{\sigma}(\mathbf{k}_\parallel,E) -\mathbf{\Sigma}_{R}^\sigma(\mathbf{k}_\parallel,E)$,
in which the self-energies act as external energy-, momentum- and spin-dependent 
potentials. In Eq.~(\ref{landauer}), 
$\mathbf{\Gamma}_{L(R)}^\mathbf{\sigma}(\mathbf{k}_\parallel,E)=i\big[\mathbf{\Sigma}_{L(R)}^\sigma(\mathbf{k}_\parallel,E)-\mathbf{\Sigma}_{L(R)}^{\sigma\dagger}(\mathbf{k}_\parallel,E)\big]$ 
is the so-called left (right) broadening matrix that accounts for the 
hybridization-induced broadening of the single-particle energy levels of 
the scattering region. Importantly, for non-interacting electrons, it has 
been proved that the Landauer and the Kubo approaches are 
equivalent~\cite{fi.pa.81}, 
so that the linear-response transport properties of a system can be computed 
with either formalism. During the last few years, the Landauer 
approach has been systematically applied in conjunction with DFT in order 
to perform calculations of the conductance of different classes of real 
nano-devices~\cite{ta.gu.01}. In this combination the DFT provides 
a single-particle 
theory in which the Kohn-Sham eigenstates are interpreted as single-particle 
excitations. Although this approach is only valid 
approximatively,
DFT-based transport studies have provided insightful results concerning the 
role of the band-structure in the electron transport process through layered 
heterostructures \cite{sc.ke.95, sc.ke.98,bu.zh.01,ru.mr.09,ca.ar.12}.

With the effect of the electron-electron interaction beyond the DFT explicitly 
considered, the retarded Green's function of Eq.~(\ref{Green_add}) is replaced by 
the following one, carrying the subscript ``MB'' for ``many-body'':
\begin{equation}
\mathbf{G}^\sigma_{MB}(\mathbf{k}_\parallel,E) = \left[
\epsilon^+\mathbf{S}(\mathbf{k}_\parallel) - \mathbf{H}^\sigma(\mathbf{k}_\parallel)
-\mathbf{\Sigma}_{L}^{\sigma}(\mathbf{k}_\parallel,E) 
-\mathbf{\Sigma}_{R}^{\sigma}(\mathbf{k}_\parallel,E) 
-\mathbf{\Sigma}_{MB}^{\sigma}(\mathbf{k}_\parallel,E)
\right]^{-1}.
\label{Green_add_MB}
\end{equation}
Here, $\mathbf{\Sigma}^\sigma_{MB}(\mathbf{k}_\parallel,E)$ is the many-body 
self-energy defined through the Dyson equation 
$\mathbf{\Sigma}^{\sigma}_{MB}(\mathbf{k}_\parallel,E)=\mathbf{G}^\sigma(\mathbf{k}_\parallel,E)^{-1}-\mathbf{G}^\sigma_{MB}(\mathbf{k}_\parallel,E)^{-1}$
\cite{mahan}. This accounts for all electron-electron interaction effects 
neglected in $\mathbf{G}^\sigma(\mathbf{k}_\parallel,E)$.
The self-energy acts as a spin-, momentum- and energy-dependent potential, 
whose imaginary part produces a broadening of the single-particle states due to 
finite electron-electron lifetime.
 In this work, the many-body self-energy is computed at the DMFT level, meaning that $\mathbf{\Sigma}^\sigma_{MB}(\mathbf{k}_\parallel,E)$ is approximated by a ${\bf k}$-independent quantity $\mathbf{\Sigma}^\sigma_{MB}(E)$, i.e.,
a spatially local but energy-dependent potential. 
Then, as suggested by Jacob {\it et al.} ~\cite{ja.ha.09,ja.ko.10}, 
the conductance and the transmission probability are obtained within the 
Landauer approach by using Eqs. (\ref{conductance}) and (\ref{landauer}), 
where one replaces  $\mathbf{G}^\sigma(\mathbf{k}_\parallel,E)$ by  
$\mathbf{G}_\mathrm{MB}^\sigma(\mathbf{k}_\parallel,E)$. This is an 
approximation, since it neglects vertex corrections due to in-scattering 
processes~\cite{oguri.01,me.wi.92}, which in general increases 
the conductivity. But we are not aware of any established method to compute 
those vertex corrections to linear-response transport within the considered framework. 
In our approach the Landauer transmission is calculated using the improved DMFT electronic structure, rather than the DFT one.
Note that the DMFT provides the single-particle excitations of the system, 
whereas the Kohn-Sham DFT eigenvalues formally do not reveal such quasiparticle 
states.

\subsection{DMFT-based computational approach}
\label{DMFT-calc}
The transport calculations are performed according to the Green's function 
scheme presented above by using the DFT-based transport code 
{\sc smeagol}~\cite{ro.su.05,ro.su.06,ru.sa.08}. The many-body self-energy 
entering in Eq. (\ref{Green_add_MB}) is calculated using the EMTO-DMFT 
method \cite{ch.vi.03,an.sa.00,vi.sk.00,vito.01} within a screened 
KKR~\cite{wein.90} approach. In both codes the 
Perdew-Burke-Ernzerhof (PBE) GGA \cite{PBE-96} for the exchange-correlation 
density functional is used. Self-consistent DFT calculations 
are performed separately in {\sc smeagol} and in the EMTO-code. The 
many-body self-energy is then evaluated after self-consistency in the 
EMTO-code, and passed to the {\sc smeagol} Green's function to compute 
the transmission along Eq.~(\ref{landauer}). The {\sc smeagol} imports the DFT 
Hamiltonian from the {\sc siesta} code~\cite{so.ar.02}, which uses 
pseudopotentials and expands the wave functions of valence electrons  
over the basis of numerical atomic orbitals (NAOs). The EMTO code, in its turn,
uses the muffin-tin construction; 
we present a detailed description of the projection of quantities such as the 
many-body self-energy from the EMTO basis set into the NAO basis set 
({\sc smeagol}/{\sc siesta}) in the Appendix.

For the EMTO-DMFT calculations, the following multi-orbital on-site 
interaction term is added to the GGA Hamiltonian in the EMTO basis:
$\frac{1}{2}\sum_{{i \{m, \sigma \} }} U_{mm'm''m'''}
c^{\dag}_{im\sigma}c^{\dag}_{im'\sigma'}c_{im'''\sigma'}c_{im''\sigma} $.
Here, $c_{im\sigma}$($c^\dagger_{im\sigma}$) destroys (creates) an electron 
with spin $\sigma$ on orbital $m$ 
{at the}
site $i$. The Coulomb matrix elements 
$U_{mm'm''m'''}$  are expressed in the standard way~\cite{im.fu.98} in terms 
of three Kanamori parameters $U$, $U'$ and $J$. Then, within DMFT the 
many-body system is mapped onto a multi-orbital quantum impurity problem, 
which corresponds to a set of local degrees of freedom connected to a bath 
and obeys a self-consistently condition~\cite{ge.ko.96,ko.vo.04}.
In the present work the impurity problem is solved with a spin-polarized 
$T$-matrix fluctuation exchange (SPTF) method~\cite{li.ka.98,ka.li.02}. 
This method was first proposed by Bickers and Scalapino~\cite{bi.sc.89} in the 
context of lattice models. In practice, it is a perturbative expansion of the 
self-energy in powers of $U$, with a resummation of a specific classes of 
diagrams, such as ring diagrams and ladder diagrams. The expansion remains 
reliable when the strength of interaction $U$ is smaller than the bandwidth 
of the bath, which is fulfilled in the case of Cu-Co-Cu heterostructures 
and for the considered values of the Coulomb parameters. The impurity 
solver we use is multi-orbital, fully rotationally invariant, and moreover 
computationally fast, since it involves matrix operations like inversions 
and multiplications. The perturbation theory can be performed 
either self-consistently, in terms of the fully dressed Green's function,
or non-self-consistently, as was done in the initial 
implementations~\cite{ch.vi.03,mi.ch.05}. When the interaction is 
small with respect to the bandwidth,
no appreciable difference exists between 
the non-self-consistent and self-consistent results~\cite{dr.ja.05,ko.sa.06}. 
This has to be distinguished from the DMFT self-consistency, which is employed
in both cases. In the present calculation,
we use non-dressed Green's functions to perform these infinite summation of diagrams. 
Moreover, we consider a different treatment of 
particle-hole (PH) and particle-particle (PP) channels. The particle-particle 
(PP) channel is described by the $T$-matrix approach \cite{gali.58} which yields 
renormalization of the effective interaction. This effective interaction is 
used explicitly in the particle-hole channel; details of this scheme can be 
found in Ref.~\cite{ka.li.02}. The particle-particle contribution to the 
self-energy  is combined with the Hartree-Fock and the second-order 
contributions \cite{ka.li.99}.
The many-body self-energy is computed at Matsubara frequencies  
$\omega_n=(2n+1)\pi/\beta$, where $n=0,1,2,...$ 
and $\beta$ is the inverse temperature. The Pad\'e~\cite{vi.se.77} analytical 
continuation is employed to map the self-energies from the Matsubara 
frequencies onto real energies, as required in the transmission calculation. 
Note that since some parts of the correlation effects are already included 
in the GGA, the double counting of some terms has to be corrected. To 
this end, we start with the GGA electronic structure and replace the obtained 
$\mathbf{\Sigma}^{\sigma}_{MB}(E)$ by 
$\mathbf{\Sigma}_{MB}^{\sigma}(E)-\mathbf{\Sigma}_{MB}^{\sigma}(0)$ 
in all equations of the GGA+DMFT method \cite{li.ka.01}, the energy $E$ here 
being relative to the Fermi energy. This is a common double-counting 
correction for treating metals; a more detailed description 
can be found in~\cite{pe.ma.03}.

\subsection{Cu-Co-Cu heterostructure setup}\label{geom}

The basis set used in the {\sc siesta} and {\sc smeagol} calculations is of 
``double-zeta with polarization'' (DZP) quality. 
In the ``standard'' {\sc Siesta} basis construction 
algorithm, there is an ``energy shift'' parameter which allows to control
the extent of basis functions on different atoms in a multi-element system
in a balanced way; in our case this parameter was taken to be 350~meV,
resulting in basis functions extending to 6.17~$a_0$ (Cu$4s$),
3.39~$a_0$ (Cu$3d$), and 6.31~$a_0$. (Co$4s$). 
For Co$3d$ states, a smaller basis function localization
was intentionally imposed, corresponding to the extention of 4.61~$a_0$,
where $a_0$ is the Bohr radius.
The basis functions are usually freely chosen and not subject to
optimization; however, in view of their quite restricted number in {\sc Siesta},
it often makes sense to look at resulting ground-state properties
of materials as a benchmark for the validity, or sufficiency,
of a basis. With the above settings, the relaxed lattice parameters of pure 
constituents were found to be $a$ = 3.65~{\AA} (fcc Cu, 1\% larger than the 
experimental value) and $a$ = 2.52~{\AA}, $c$ = 4.06~{\AA} (hcp Co, both within 0.5\%
of experimental values). Moreover the magnetic moment per Co atom was
1.65~$\mu_{\rm B}$ (%1.65~$\mu_{\rm B}$ in 
equal to experiment).

\begin{figure}[h]
\includegraphics[width=0.75\columnwidth]{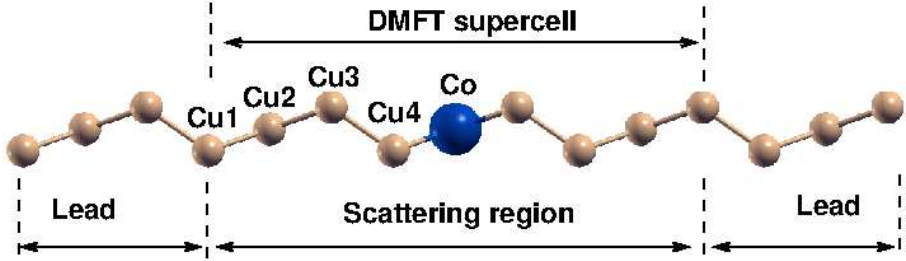}
\caption{(color online) Schematic representation of the supercell used in 
calculations. The length of the cell and the distances between Cu1, Cu2, Cu3,
Cu4 and Co are given in the text. The unit cell dimensions of the leads are 
kept at the relaxed bulk value 3.65 {\AA}.
}
\label{fig1}
\end{figure}

In order to calculate the transport properties, semi-infinite leads 
are attached on both sides of the scattering region. We consider Cu(111)-cut
leads, characterized by the ABCABC atomic plane repetition 
into the leads, i.e., along the transport direction henceforth referred to
as $z$. It is assumed that the ABCABC layer sequence is smoothly
continued throughout the scattering region, including the Co monolayer
(see Fig.~\ref{fig1}). To model the scattering region
the sequence is repeated and the Co layer is considered to replace the Cu layer.

In {\sc smeagol} the Hamiltonian of the scattering region is 
matched to that of the leads at the boundary of the scattering region,
thus implying that whatever perturbation is induced by a scatterer it has 
to be confined witin the scattering region. 
In other words,
the simulation cell needs to contain enough Cu layers on each 
side of the Co layer
to ``screen'' it completely.
We verified that using seven layers on each side provides a good agreement
between the potential at the boundaries of our setup with the one from the 
periodic Cu leads calculation. The resulting cell geometry is shown in 
Fig.~\ref{fig1}: the ``Lead''+``Scattering region''+``Lead'' composes the 
{\sc smeagol} cell, merging on its two ends with  the unperturbed 
semi-infinite Cu electrodes. A restricted spatial relaxation within the 
scattering region was done by {\sc Siesta}, whereby the total thickness 
of this region varied in small steps, and the $z$-positions of Cu2, Cu3 
and Cu4 layers between the limiting Cu1 and the central Co layer were 
adjusted till the forces fell below 0.01~eV/{\AA}. This resulted in 
interlayer distances of 2.119~{\AA} (Cu1-Cu2 and Cu2-Cu3), 2.118~{\AA} 
(Cu3-Cu4) and 2.104~{\AA} (Cu4-Co). The length for the supercell along 
the $z$-direction resulting from the minimization of the cell total
energy was 31.619~{\AA}.
The relaxed structural parameters obtained within the GGA 
were then used in the GGA+DMFT calculation, and no additional 
structure relaxation was attempted at the GGA+DMFT level.

\section{Results}
\label{results}
In this section we discuss the changes in the electronic structure and in 
the conductance of a single Co layer sandwiched between semi-infinite 
Cu electrodes caused by the inclusion of the Coulomb interaction at the 
GGA+DMFT level. The chosen values for Coulomb and exchange parameters for 
the $3d$-Co orbitals are $U=3$~eV and $J=0.9$~eV, while no interaction 
beyond GGA is considered for the $3d$-Cu states neither in the scattering 
region nor in the leads. The values of $U$ and $J$ are sometimes used as 
fitting parameters, although it is possible, in principle, to compute the 
dynamic electron-electron interaction matrix elements with good 
accuracy~\cite{ar.im.04}. The static limit of the energy-dependent
screened 
Coulomb interaction leads to a $U$ parameter in the energy range between 2 
and 4 eV for all $3d$ transition metals, with substantial variations related 
to the choice of the local orbitals \cite{mi.ar.08}. As the $J$ parameter 
is not affected by screening it can be calculated directly within LSDA; 
it turns out to be about the same for all $3d$ elements, $J$ 
$\approx$ 0.9 eV~\cite{im.fu.98}. The sensitivity of results to
$U$ and $J$ will be briefly addressed towards the end of this section.
As regards the temperature, two values $T=80$ K and $200$ K are addressed.
Smaller temperatures could be considered; however, this 
would strongly increase the computational efforts connected with the analytical 
continuation of the data onto the real axis.

\subsection{Electronic structure calculations}
\begin{figure}[h]
\includegraphics[width=0.5\columnwidth,clip=true]{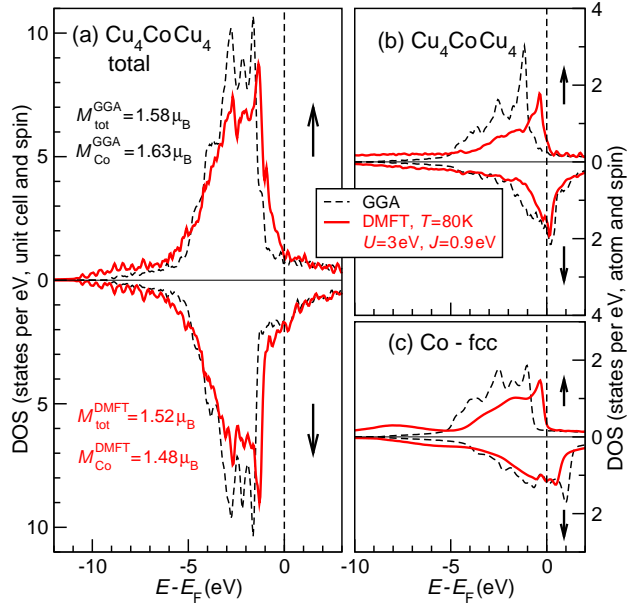}
\caption{\small (Color online) 
Densities of states calculated by EMTO in relaxed geometries. Dashed black 
lines: GGA results; solid red lines: GGA+DMFT results. (a) Total DOS per 
scattered region; (b) local DOS per central Co atom
of the heterostructure; (c) local DOS per atomic sphere of the same size
in pure fcc bulk cobalt. The values of total and local Co magnetic moments
for the scattered region are indicated in (a).
}
\label{fig2}
\end{figure}

The total density of states (DOS) for the Cu-Co-Cu heterostructure is shown 
in Fig.~\ref{fig2}(a). The Co contribution to the total DOS,
attributed to the atomic sphere radius of 2.69~$a_0$, is presented 
in Fig.~\ref{fig2}(b). For comparison we plot in Fig.~\ref{fig2}(c) 
the Co DOS in the bulk fcc structure,
with the same atomic sphere radius.

We start with discussing the features of the electronic structure of 
bulk Co. For the majority-spin electrons, the GGA DOS (Fig.~\ref{fig2}c) 
is fully occupied. In the minority spin-channel, the Fermi level
falls between two pronounced peaks at $\sim E_F \pm$1~eV.
The orbital occupations are shown in Tab.~\ref{tab1}. According to the
GGA results the majority spin-up channel has a nominal $d$-occupation of 
$4.70$ while for minority electrons the occupation amounts to $2.87$. 
The $s$-electrons carry a negligible polarization, while $p$-electrons are slightly 
spin-polarized with a sign  opposite to the main $d$-polarization which 
establishes a magnetic moment of $1.74~\mu_B$. 
As a consequence of the local Coulomb interactions parameterized 
by $U$ and $J$ within DMFT, the DOS distribution changes considerably. 
The overall broadening is strongly modified by the imaginary part of the 
self-energy. The top of the occupied $d$-band in the majority spin channel 
is shifted closer to the Fermi level, and some redistribution of the spectral 
weight occurs. These changes do not noticeably affect the occupation of 
$s$-orbitals, however, the magnetic moment, mostly due to $d$ electrons,
is significantly reduced from $1.74~\mu_{\rm B}$ to $1.41~\mu_{\rm B}$.

\begin{table}%[h]
\caption{Orbital occupations and magnetic moments for the Cu-Co-Cu
heterostructure and bulk Co-fcc. The DMFT calculations have been
performed for $T$ = 200~K, $U$ = 3~eV, $J$ = 0.9~eV.\label{tab1}
}\medskip
\begin{tabular}{ccccccccc}
\hline
 & \multicolumn{3}{c}{$n^{\rm GGA}$} & $M^{\rm GGA}$ & 
 \multicolumn{3}{c}{$n^{\rm DMFT}$} & $M^{\rm DMFT}$  \\
 \cline {2-4} \cline{6-8}
 Atom & $s(\downarrow\!/\!\uparrow)$ & 
        $p(\downarrow\!/\!\uparrow)$ & 
        $d(\downarrow\!/\!\uparrow)$ & ($\mu_{\rm B}$) &
        $s(\downarrow\!/\!\uparrow)$ & 
        $p(\downarrow\!/\!\uparrow)$ &
        $d(\downarrow\!/\!\uparrow)$ & ($\mu_{\rm B}$) \\
\hline
\multicolumn{9}{c}{Co bulk-fcc:} \\
 Co: & (0.34/0.33) & (0.39/0.31) & (2.87/4.70) & 1.74
     & (0.34/0.33) & (0.38/0.35) & (3.04/4.50) & 1.41 \\
\hline
\multicolumn{9}{c}{Cu$_4$CoCu$_4$ scattering region:} \\
 Cu1-3:  & (0.36/0.36) & (0.33/0.33) & (4.78/4.78) & 0.00 
         & (0.39/0.39) & (0.45/0.45) & (4.72/4.72) & 0.00 \\
 Cu4:    & (0.36/0.36) & (0.36/0.33) & (4.76/4.79) & 0.00
         & (0.38/0.37) & (0.42/0.39) & (4.58/4.65) & 0.02 \\
 Co:     & (0.33/0.32) & (0.33/0.31) & (2.96/4.63) & 1.63
         & (0.32/0.33) & (0.33/0.37) & (2.70/4.13) & 1.48 \\
\hline
\end{tabular}
\end{table}

Essentially the correlation effects are 
determined not only by the magnitude of the local Coulomb parameters ($U,J$), 
but also by the orbital occupations. It was argued~\cite{mo.ma.02,gr.ma.07} 
that electronic interactions may lead to the creation of either a majority-spin or 
a minority-spin hole. As the majority-spin channel is essentially full, 
there is effectively no space for excitations just across the Fermi level. 
On the contrary, in the minority-spin channel one finds a high density of 
electrons which can be immediately excited, leaving back holes.
Such an occupation asymmetry has consequences concerning possible interaction 
channels in the multi-orbital Hubbard model: A majority-spin hole can only 
scatter with opposite-spin particles, which would cost an effective 
interaction $U$, while a 
minority-spin hole may also scatter with parallel-spin particles with 
the effective interaction $U-J<U$~\cite{ma.be.99}. Therefore correlation 
effects are expected to manifest themselves differently for majority- and 
minority-spin electrons.

Many DOS features of Co in the heterostructure geometry (Fig.~\ref{fig2}b) 
resemble those of bulk cobalt (Fig.~\ref{fig2}c), the occupation numbers 
of which are also given in Tab.~\ref{tab1}. As expected, the spin polarization 
in $s$- and $p$-channels is very small; moreover it is opposite to the 
$d$-electrons, yielding an overall magnetic moment of $1.63~\mu_B$. As 
compared to the GGA case for the central Co layer of the heterostructure, in 
GGA+DMFT the Co~$s$- and $p$-electron spin polarization changes sign, and the 
$d$-electrons spin splitting decreases. In the Cu-Co-Cu heterostructure 
geometry the Co-$d$ orbitals experience hybridization with the neighboring 
Cu-$d$ orbitals  which leads to a change in the DOS of the Cu-layer in the 
vicinity of the Co-layer.

\begin{figure}[h]
\includegraphics[width=0.5\columnwidth,clip=true]{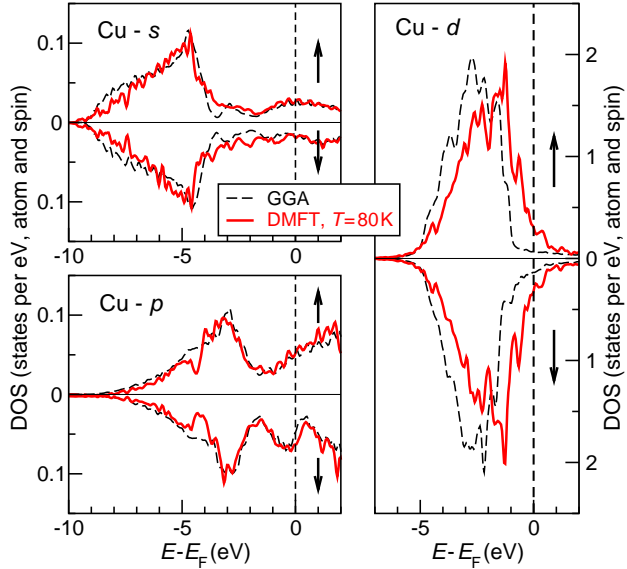}
\caption{(Color online) Orbital resolved DOS for Cu4 in the 
Cu$_4$CoCu$_4$ heterostructure computed within the GGA (black-dashed line) and 
GGA+DMFT (red solid line) for $U=3$~eV and $J=0.9$~eV.}
\label{fig3}
\end{figure}

Figure~\ref{fig3} depicts the local DOS of the Cu atom closest to 
the Co monolayer (indicated Cu4 in Table~\ref{tab1}). Even as no 
on-site interaction terms have been added to the $3d$-Cu states, the Co 
self-energy has a large impact on the GGA-DMFT density of states.
In fact, the $3d$-Cu4 states are strongly coupled with the correlated 
$3d$-Co states and are dragged towards the Fermi energy, thus increasing the 
hybridization with the $4s$- and $4p$-Cu4 states. In contrast, 
the three outmost (from Co) copper layers Cu3, Cu2, Cu1 have very similar 
orbital occupations which slightly differ from those of the Cu4.  
The inclusion of interaction in the spirit of DMFT has only a slight effect
upon the orbital occupations within Cu3, Cu2, and Cu1 as compared to 
Cu4 (see Table~\ref{tab1}). Essentially the $4s$- and $4p$-Cu4 orbitals 
slightly increase in occupation, while $3d$-orbitals are depleted accordingly. 
At the same time, the minority/majority spin 
contrast gets enhanced: from $4.76/4.79$ in GGA to $4.58/4.65$ in GGA+DMFT. 
The spectral weight transfer in the Co layer -- a consequence of electron 
correlations -- modifies slightly, through $d-d$ hybridization, the 
spin asymmetry in $d$-holes of the closest copper layer inducing a
magnetic moment. 

We note that the temperature dependence of the DOS is negligible, and the 
effectiveness of electronic correlations is not significantly different 
in the heterostructure, as compared to the case of pure-Co fcc bulk.

\begin{table}[h]
\caption{Effective mass enhancements $(m^{\ast}/m)^{\uparrow,\downarrow}$
for different bands of $d$-symmetry, calculated
according to Eq.~(\ref{Eq:eff_mass}) as a function of the
Coulomb and the exchange parameter, $U$ and $J$.\label{tab2}}
\medskip
\begin{tabular}{cc@{\hspace*{6mm}}ccc@{\hspace*{6mm}}ccc}
\hline
$U$~(eV) & $J$~(eV) &
$xy_\downarrow$& $yz_\downarrow$ & $z^2_\downarrow$ &
$xy_\uparrow$  & $yz_\uparrow$   & $z^2_\uparrow$ \\
\hline
1 & 0.3 & 1.551 & 1.703 & 1.617  & 1.578  & 1.737  & 1.656 \\
2 & 0.6 & 1.625 & 1.797 & 1.698  & 1.654  & 1.837  & 1.743 \\
3 & 0.9 & 1.661 & 1.791 & 1.703  & 1.697  & 1.856  & 1.753 \\
\hline
\end{tabular}
\end{table}

These changes are typical for correlation effects in transition metals, 
where the self-energy near the Fermi level has Fermi-liquid character:
for the imaginary part, we have 
$-\mathrm{Im}\Sigma^{\sigma}_{MB,\alpha}(E)\propto E^2$,  whereas the real 
part has negative slope, $\partial\mathrm{Re}\Sigma_{\mathrm{MB},\alpha}^{\sigma}(E)/\partial E<0$. 
Here $E$ is the energy relative to the Fermi level, and $\alpha$ 
numbers the three groups of $d$ states  in hexagonal
symmetry, $(z^2)$, $(xz,yz)$ and $(x^2-y^2,xy)$. From the self-energy 
we can also evaluate the mass enhancement \cite{mahan}, which within DMFT 
amounts to 
\begin{equation}
\left( \frac{m^*}{m_b} \right)^{\sigma}_{\alpha} = 1- 
\frac{\partial} {\partial E} {\rm Re} \ \Sigma_{\mathrm{MB}}^{\sigma,\alpha}(E),
\label{Eq:eff_mass}
\end{equation}
where $m_b$ represents the band-mass obtained within the GGA calculations. 

The values are given in Table~\ref{tab2}, and we note that the enhancement
factors for all orbitals are similar, in the range of  $1.6 - 1.8$, 
which indicates that the system is medium-correlated. 

\subsection{Transport properties}
\begin{figure}[h]
\includegraphics[width=0.45\columnwidth]{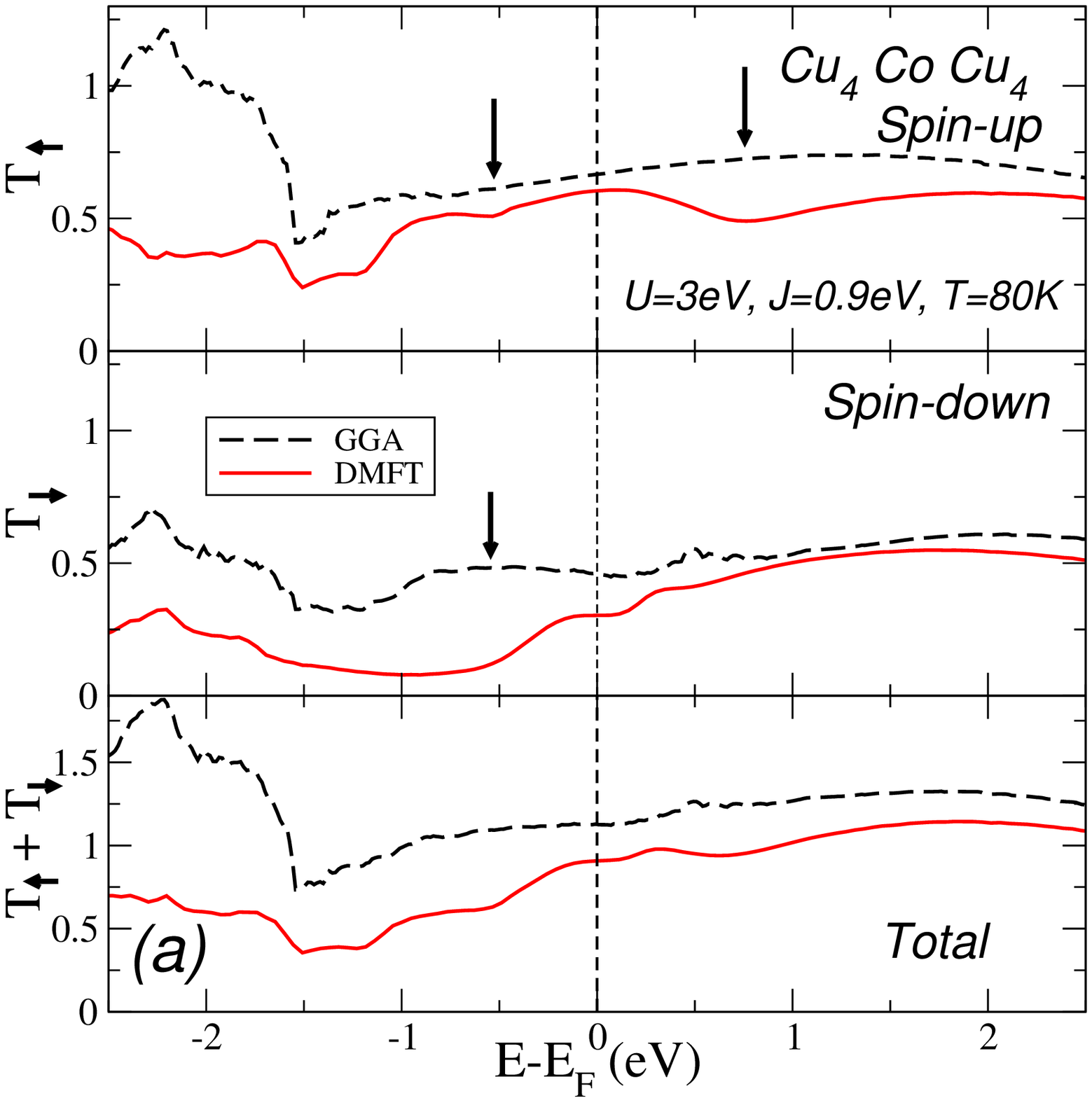}
\includegraphics[width=0.45\columnwidth]{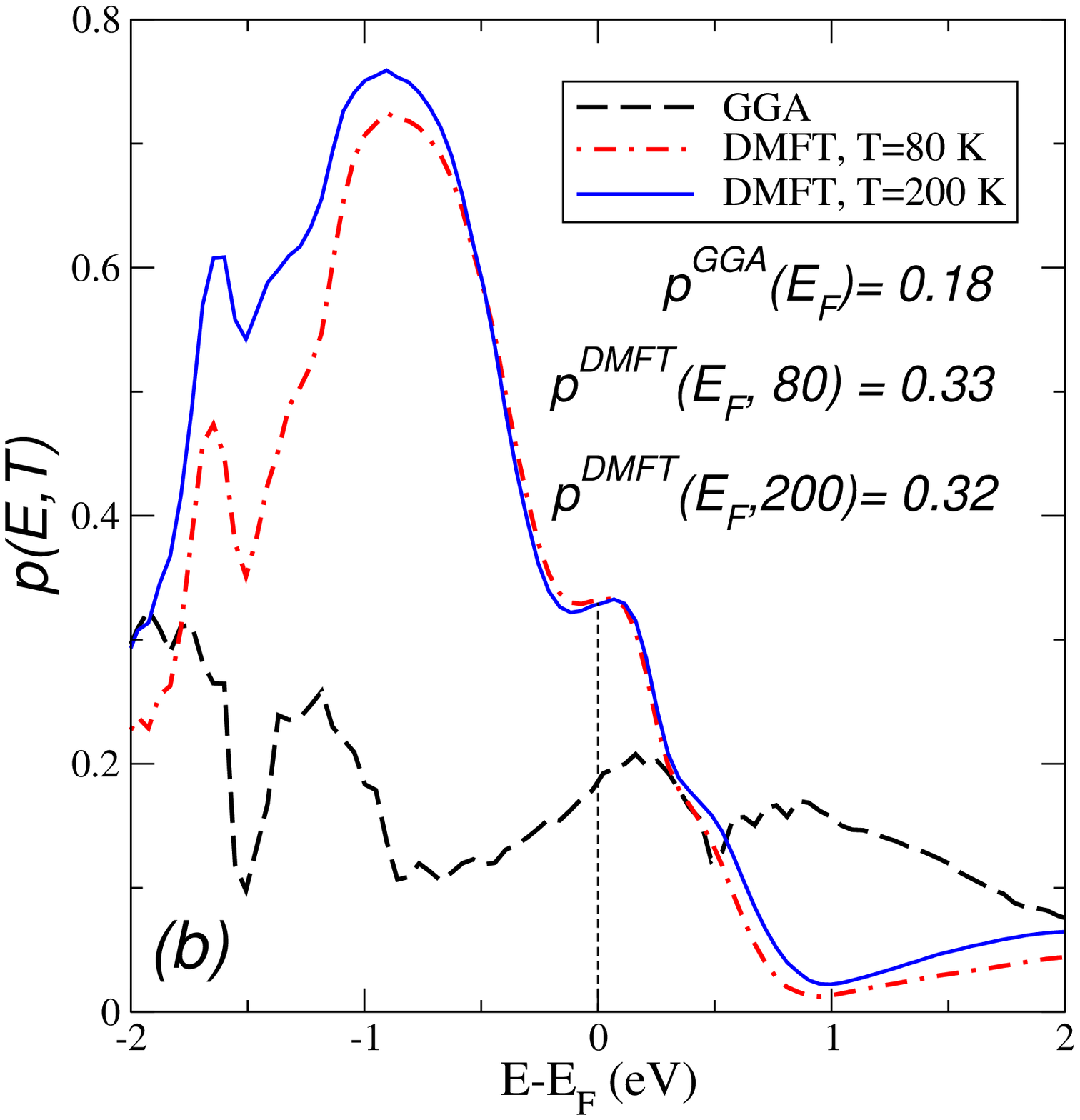}
\caption{(Color online) Left: Spin-resolved transmission: majority-spins
$T_{\uparrow}$ (upper panel), minority spins $T_{\downarrow}$ (middle panel) 
and total (lower panel). Black-dashed/red-solid lines are the GGA/GGA+DMFT 
results. Right: Transmission spin polarization around the Fermi energy 
GGA (black-dashed) and at finite temperatures T=80K (red-dotted-dashed), 
$T=200$ K(blue solid). The Coulomb and exchange
parameters are $U=3$~eV, $J=0.9$~eV. }
\label{fig4}
\end{figure}

Turning to transport properties, we display in Fig.~\ref{fig4}(a) the 
total and spin-resolved transmission probabilities computed with GGA and 
GGA+DMFT. The spin-resolved transmission probability, 
$T_\sigma(E)$, is obtained from the {\bf k}-dependent transmission, 
Eq.~(\ref{landauer}), by integrating over all $\mathbf{k}_\parallel$-points, 
so that $T_\sigma(E)=\frac{1}{\Omega_\mathrm{BZ}}\int_\mathrm{BZ}d\mathbf{k}_\parallel T_\sigma(\mathbf{k}_\parallel,E)$.
By inspecting Fig.~\ref{fig4}(a) it can be seen that the overall transmission 
is a smooth function of energy, and has a rather large value of about 
0.5 $e^2/h$ in both spin channels for most considered energies, which 
reflects the fact that we deal with an all-metal junction. 
In GGA the transport is mainly dominated by the Cu-$4s$, -$4p$ states, which
are transmitted across the Co layer passing through the Co $4s$ states, while
the Co $3d$ states do not contribute significantly to the transmission in this
energy range. The Cu $3d$ states contribute to the transmission only at
energies below $-1.5$~eV.
Note that the GGA+DMFT transmission is always smaller than the GGA one. 
The black arrows in Fig.~\ref{fig4}(a)
indicate the energies at which a significant departure between the GGA+DMFT
and the GGA transmission is observed. Specifically, the spin down transmission
drops at about $0.3$ eV below the Fermi level, where the Co $3d$ DOS is high 
in the DMFT results, while the GGA transmission stays rather
constant. In contrast, the spin-up transmission shows a ``bump'', which
extends over a region of about $2$ eV around the Fermi level. The slight dip
within this bump at about $-$0.5 eV is at the same energy as the peak in the
Co 3$d$ DOS, and represents a Fano-type reduction of transmission in such a
metallic system due to interference of electrons in different conducting
channels~\cite{ja.ha.09,ja.ko.10,ca.fe.09}.
In our calculations this feature is a consequence 
of electronic correlations on the Co atom, which through the 
$d-d$ hybridization induce spin-polarization of the $3d$-Cu4 states and 
simultaneously produce the shift in the DOS of Fig.~\ref{fig3}. 
In general, we 
note that for such all-metal systems the relation between DOS and transmission 
is non-trivial, since interference effects can lead to enhanced transmission 
also for energies with low DOS; alternatively, for high DOS the increased 
number of pathways for electrons can lead to a decrease of transmission.

From a many-body perspective, the added self-energy contributes in dephasing 
the 
electrons during the flow through the scattering region, so that the Landauer 
transmission computed with the many-body Green's function is expected to be 
reduced in comparison with the DFT case. In principle, the opposite effect, 
namely the effective in-flow of electrons from the many-body self-energy 
``electrode'' into the scattering region, would tend to increase the 
transmission. However, this in-flow process is not included in our 
calculations, as it is related to the vertex corrections~\cite{oguri.01}. 

The spin polarization of the transmission is computed according to the 
formula
\begin{equation}
p(E_{\rm F})
 = \frac{T_{\uparrow} (E_F) -T_{\downarrow} (E_F)}
{T_{\uparrow} (E_F) + T_{\downarrow } (E_F) },
\label{spin_pol}
\end{equation}
for either DFT, or DFT+DMFT, where 
$T_{\sigma=\uparrow,\downarrow}$ is the 
transmission for the spin channel $\sigma$. The spin polarization in 
transmission obtained by GGA yields $p^{\rm GGA}(E_F)=0.18$ [see also 
Fig.~\ref{fig4}(b)], while  the GGA+DMFT value reaches 0.33 at the Fermi 
level, and increases up to almost 0.8 at slightly lower energies. 
These results demostrate that electronic correlations may be decisive 
and lead to 
an increase in the spin polarization of transmission. As seen
in Fig.~\ref{fig4}(b), the enhancement in the spin polarization with
respect to the GGA result is essentially temperature independent 
in the energy range of $E_F \pm 0.5$eV. Therefore we conclude 
that the enhanced spin contrast in transmission is a many-body
effect rather than a temperature fluctuation effect. 

\begin{figure}[h]
\includegraphics[width=0.5\columnwidth]{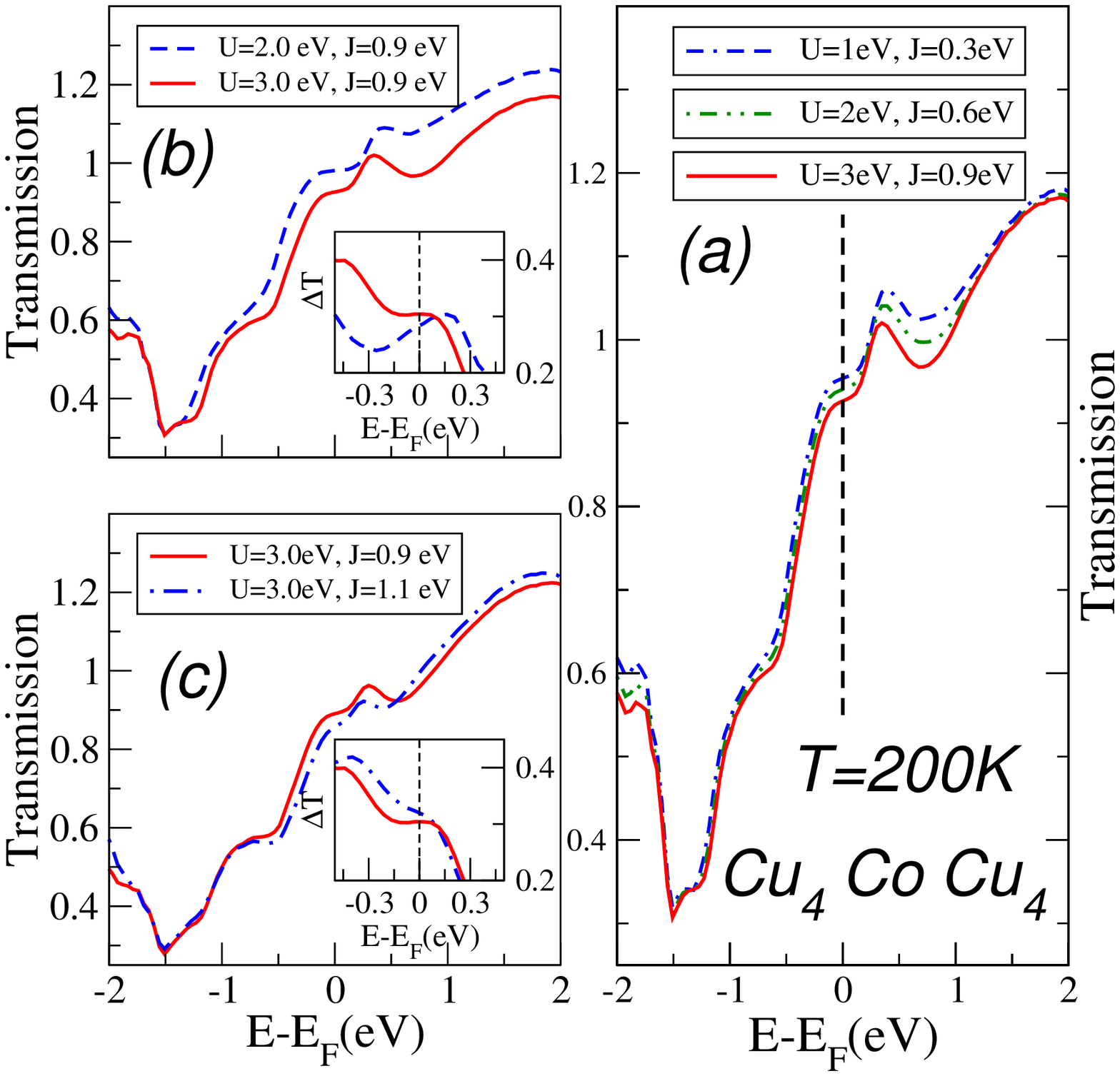}
\caption{(Color online) Transmission for different values of Coulomb
parameters: (a) with a fixed ratio $U/J$, (b) fixed $J$ and 
(c) fixed $U$, at $T=200$~K.} 
\label{fig5}
\end{figure}

%%%%%%%%%%%%%%%%%%%%%%%%%%%%%%%%%%

Finally, we test how the results for the transmission depend on the 
strength of the local Coulomb interaction parameters $U$ and $J$. 
The interaction matrix elements $U_{mm'm''m'''}$ are usually parametrized 
using Slater integrals ($F^k$) with $k=0, 2, 4$~\cite{im.fu.98}. 
Accordingly the Hubbard $U$ parameter is constructed as a simple average 
over all possible pairs of correlated
orbitals and is identified with the Slater integral $U=F^0$.  The other 
Slater integrals $F^2, F^4$ are fitted to the multiplet structure measured 
in X-ray photoemision~\cite{an.gu.91}. An empiric relation has been 
introduced which connects the magnitude of the second and forth order
Slater integrals, $F^4/F^2 \approx 0.625$~\cite{gr.fu.90}. The 
Hund's exchange $J$ 
is expressed in terms of $F^2$ and $F^4$ which for the $d$-shell takes
the form $J=(F^2+F^4)/14$~\cite{an.so.93}, therefore the knowledge of 
the $(U,J)$-pair allows to compute all the matrix elements $U_{mm'm''m'''}$.
In Fig.~\ref{fig5}(a) we plot the transmission (computed at $200$ K) 
keeping the ratio $U/J=1.0/0.3$ constant. While scaling the ratio 
$\alpha \cdot U/J$ with $\alpha=1,2,3$ we observe a monothonic reduction 
of the transmission at the Fermi level. This result is expected, as the 
matrix elements of the interaction are scalled in magnitude. 
In the same time, larger mass enhancement factors are obtained as $\alpha$
increases (see Tab.~\ref{tab1}).
Consequently we may conclude that the heavier the electron is, 
the smaller is the transmission at the Fermi level.
Within $\pm 0.15$~eV of the Fermi level, a flat region in the transmission
can be seen. 
Beyond these ranges, we note that below the Fermi level,
down to $-$1~eV from it, the transmission decreases almost indiscriminately
for different values of $(U,J)$. In the positive energy range up to roughly 
$+1$~eV, on the contrary, the transmission values differ, with larger $(U,J)$
resulting in lower transmission.

In Fig. \ref{fig5}(b) we display the dependence on parameters differently 
keeping $J=0.9,$~eV constant and varying $U$.
We note that with higher $U$, the flat region centered at 
$E_{\rm F}$ shrinks a bit, which can be traced back 
to a stronger presence of $d$ orbitals in the correlated transmission. 
The inset of Fig. \ref{fig5}(b) depicts the ``contrast'' or spin difference in 
transmission: $\Delta T(E) =T_\uparrow(E) -T_\downarrow(E)$. 
The spin contrast changes slope
as $U$ varies, from markedly positive $d(\Delta T)/dE$ at 
$E_{\rm F}$ for $U=2$~eV to roughly zero for $U=3$~eV.

In Fig.~\ref{fig5}(c), on the contrary, the $U$ parameter is kept fixed 
to a ``good'' (yielding large flat region) value $U=3$~eV, and two different 
values are taken for the $J$ parameter. On increasing $J$, the flat region 
around $E_{\rm F}$ gets further reduced. Simultaneously (as seen in the 
inset and contrary to the behavior depicted in Fig.~\ref{fig5}b),
$\Delta T(E)$ acquires a negative slope. The change in the slope is 
potentially an interesting effect to be exploited in thermoelectric 
transport.   

\section{Conclusion}
\label{conclusion}

In this work we studied the effects of local electronic interactions and 
finite temperatures upon the transmission across the Cu$_4$CoCu$_4$ metallic
heterostructure. Electronic structure calculations were performed using 
GGA and GGA+DMFT, assuming a local Coulomb interaction in the Co layer. 
We used a fully rotationally invariant Coulomb interaction on cobalt 
$d$-orbitals. The effective mass enhancement ratio for all orbitals is in 
the range of $1.6$ to $1.8$, suggesting that we deal with medium correlated 
system. Concerning the density of states, the presence of Coulomb interactions 
leads to a shift of the majority-spin channel of Co $d$-orbitals towards the 
Fermi level and to 
a redistribution of the spectral weights. In the minority spin-channel, 
the changes are less pronounced. This difference leads to different 
correlation effects for the majority- and minority-spin electrons.
All these
causes a decrease of the overall Co magnetic moment (with a predominant 
$d$ character) from $1.63~\mu_B$ (GGA)to $1.48~\mu_B$ (GGA+DMFT).

The transmission probability
has been computed combining two different ab-initio codes, 
EMTO and {\sc Siesta}/{\sc Smeagol}. In order to transfer the 
many-body self-energy computed within the EMTO code into 
{\sc Siesta}/{\sc Smeagol}, we used a nearly unitary transformation which 
can be determined by requiring that the expectation value of the occupation 
matrix should be representation independent. The methodology was illustrated 
for a two-orbitals model (see  Appendix~\ref{basis_transf}) and was 
carried out numerically for the Cu$_4$CoCu$_4$ heterostructure. Note that 
such a transformation is rather general and can be used in transfering
quantities 
between two different implementations. Several tests confirmed that the 
proposed method is robust and numerically stable. 
With this combination of methods we have studied the transmission as a 
function of temperature and Coulomb parameters, which reveals the metallic 
character of the system considered. Substantial differences in the  
conducting processes, related to the presence of local 
Coulomb interaction in the cobalt layer were observed, due to changes in 
the electronic structure. Generally the transmission 
decreases with interaction, although the relation between changes in the 
electronic configuration and the transmission is highly non-trivial, due 
to interference effects. With electron correlations properly taken into 
account, the total transmission at the Fermi level drops by about $20\%$,
whereas its spin polarization (spin contrast) increases 
by about $40\%$. These effects are entirely a consequence of the electronic 
correlation, since the transmission is practically temperature independent 
in the range of $E_F \pm 0.5$~eV. This suggest that the enhanced spin 
contrast in transmission is fully a many-body effect. 
In order to quantify the spin polarization effects in the transmission, 
we studied the transmission difference $\Delta T(E)$.  This quantity 
clearly displays a strong dependence on the Coulomb parameters.

In conclusion, we have shown that electronic correlations may considerably 
affect the transmission and spin filter properties of heterostructures, even 
though the correlations would be classified only as ``medium'' when considering 
the effective mass enhancement. Hence results based on studies {\it neglecting} 
electronic correlations, which are numerous, must be interpreted with caution.

\section*{Acknowledgement}
The calculations were performed in the Data Center of NIRDIMT. Financial
support offered by the Augsburg Center for Innovative Technologies, and 
by the Deutsche Forschungsgemeinschaft (through TRR 80) is gratefully 
acknowledged. A.D. and I.R. acknowledge financial support 
from the European Union through the EU FP7 program through project 
618082 ACMOL. M.R. acknowledges the Support by the Serbian Ministry of Education, 
Science and Technological Development under Projects ON171017 and III45018.
A.\"{O}. would like to acknowledge financial support from 
the Axel Hultgren foundation, from the Swedish steel producer's 
association (Jernkontoret) and the hospitality received at the 
Center for Electronic Correlations and Magnetism, Institute of 
Physics, University of Augsburg, Germany. L.V. acknowledges 
support from the Swedish Research Council.

\section*{APPENDIX}
\appendix

\section{SIESTA and EMTO density of states}
\begin{figure}[h]
\includegraphics[width=0.7\columnwidth]{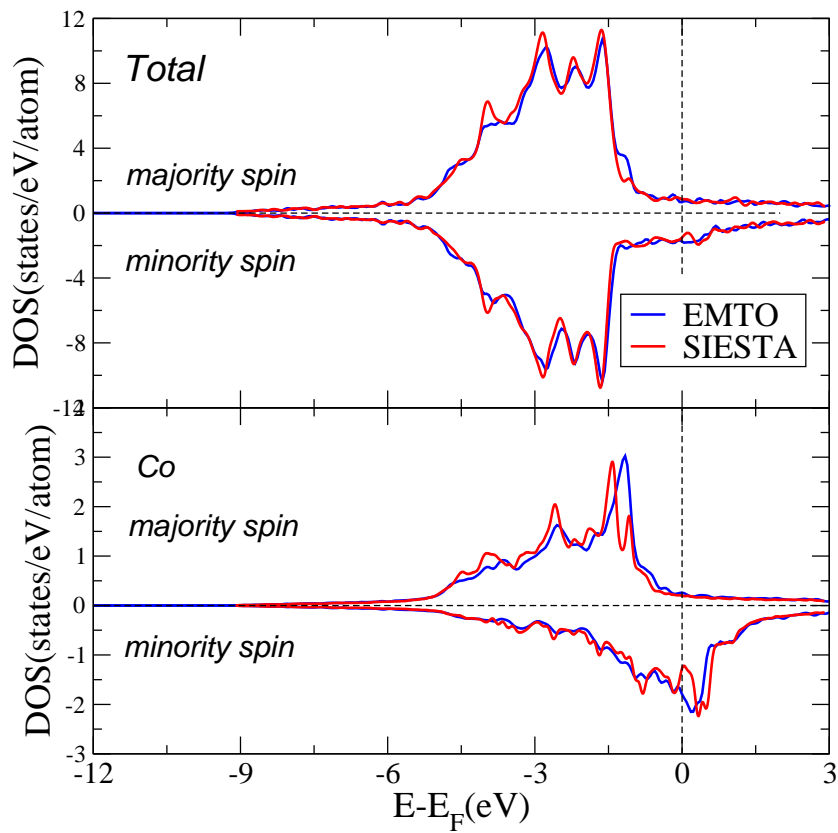}
\caption{(color online) Total (unit cell) and Co density of states
computed within the {\sc Siesta} and EMTO technique.}
\label{fig6}
\end{figure}

Inorder to demonstrate the reliability of both codes 
concerning the electronic 
structures, we present below the density of states for the 
heterostructure (Fig. \ref{fig6}). 
A rather good agreement is apparent.

\section{Matrix elements of the self-energy in the NAO  basis set}
\label{basis_transf}
The matrix representation of the self-energy operator is determined by the 
chosen basis. Within the EMTO basis set it has the form 
$\delta_{RR'}{\tilde \Sigma}_{RL,RL'}^{\sigma}(z)$. Here $R$ and $R^\prime$ 
are site indices while the $L$ symbol labels the orbital quantum numbers. To 
compute the transmission/conductance it is desirable to work within the 
{\sc Siesta}/{\sc Smeagol} basis. We emphasize that the major part of 
calculation is done within the {\sc Siesta}+{\sc Smeagol} package. 

Since significant methodological differences exists 
between {\sc Siesta} and EMTO, in the following we discuss the 
methodology of data transfer between the two codes. We describe briefly 
the most significant differences. 
The former ({\sc Siesta}) implementation uses norm-conserving pseudopotentials,
whereas the latter (EMTO) uses an all-electron formulation. {\sc Siesta} uses 
no shape approximation with respect to the potential, whereas EMTO relies on 
the 
muffin-tin concept~\cite{an.sa.00,vito.01}. Basis functions in {\sc Siesta} 
are atom-centered numerical functions, whose angular parts are spherical 
harmonics, and radial parts are strictly confined numerical functions. 
A tradition finding its origins in quantum chemistry suggests that, in 
order to improve variational freedom of the basis set, more than one 
radial function is adopted into the basis for a given angular combination 
$(l,m)$, referred to as ``multiple-$\zeta$'' basis orbitals.

Even as {\sc Siesta} maintains flexibility in constructing the basis set 
out of different ``zetas'', possibly including moreover 
``polarization orbitals'' and allowing a variety of schemes to enforce 
confinement, we shall stick in the following to the case of just
``double-$\zeta$'', i.e., $2{\times}5=10$ basis functions, provided to 
accurately describe the $3d$ states on each cobalt atom. We shall fix some 
notation for further reference. The cumulate index of a basis function will 
be $\mu{\equiv}\{Ilm\zeta\}$, where $I$ indicates the atom carrying the basis 
function, $\zeta$ numbers the ``zeta''s ($=1$ or $2$ in our case),
and the $(l,m)$ is the conventional angular moment index. It should be 
noted, however, that {\sc Siesta} employs real combinations 
of ``standard'' spherical harmonics, so that the indices $m={-2}$ through $2$ 
for $l=2$ correspond to $xy$, $yz$, $3z^2\!-\!r^2$, $xz$  and
$x^2\!-\!y^2$ $d$-functions, correspondingly. With the above notation, the 
$i$th eigenstate of the Kohn--Sham Hamiltonian will expand into the basis 
functions $\phi_{\mu}$ as follows:
\begin{equation}
\Psi_i({\bf r})=\sum\limits_{\mu}c_{\mu i}\,\phi_{\mu}({\bf r}-{\bf R}_I)\,,
\end{equation}
and the variational principle yields the expansion coefficients \cite{so.ar.02}.
Within the EMTO, the $d$-orbitals manifold is constructed using a basis set 
with real harmonics (physical orbitals representation), and the occupation 
matrix is obtained integrating the complex contour Green's function (properly 
normalized path operator) in terms of the exact muffin-tin orbitals 
$\Psi_{RL}(\epsilon, {\bf r})$ corresponding to the energy $\epsilon$ 
\cite{an.sa.00,vi.sk.00,vito.01}.  
On the other hand, in {\sc Siesta} the numerical basis set is not restricted to 
the physical orbitals and allows the definition of simple/double $\zeta$
``atomic-like'' representations. Enhancing the numerical atomic orbitals  
basis vectors does not affect the dimension of the vector spaces (in both 
cases the $d$-subspace), however, it complicates the algebra for the 
transformation matrix.
The transformation involves the double zeta basis set,
$\phi_{\mu}$,  which for the $L(l,m)$-manifold contains $2(2l+1)$ 
components splitted into the first $(2l+1)-$single-zeta and the 
second $(2l+1)-$double-zeta components. First we write the explicit form 
for the EMTO orbital in a vector form for the $L(l=2,m=2l+1)$ subspace 
$|  \Psi_{RL}(\epsilon, {\bf r})  \rangle $, and 
the corresponding {\sc Siesta} basis vector in the double $\zeta$-basis,
$|  \phi_{RL}( {\bf r})  \rangle =
(\phi^{1\zeta}({\bf r}) , 
\phi^{2\zeta}({\bf r}) ) $.
The general transformation  matrix $V_{m_i,m_j}^{1 \zeta, 2 \zeta} (\epsilon)$
is defined as the inner products of 
$\Psi_{R^\prime L^\prime}(\epsilon, {\bf r})$
with {\sc Siesta}'s double-$\zeta$ basis, and takes the form of a dyadic 
product:  
\begin{eqnarray}\label{kdot1}
V_{m_i,m_j}^{1 \zeta, 2 \zeta} (\epsilon) &=&   \langle \Psi_{R^\prime L^\prime}(\epsilon, {\bf r}) | \otimes 
|  \phi_{RL}( {\bf r}) \rangle = |  \Psi_{RL}(\epsilon, {\bf r})  \rangle^T |  \phi_{RL}( {\bf r})  \rangle   \\
& =& 
\begin{pmatrix}
V_{m_i,m_j}^{1\zeta} (\epsilon)  \\
%\hline
V_{m_i,m_j}^{2\zeta} (\epsilon) 
\end{pmatrix} \nonumber 
\end{eqnarray}

The definition Eq. (\ref{kdot1}) for the $V$ matrix suggests the 
possibility of an explicit construction. However, a couple of obvious 
ambiguities may arise: (i) the transformation carries an energy dependence 
originating from the energy dependence of the EMTO orbitals 
\cite{an.sa.00,vi.sk.00,vito.01}, in contrast to
the NAO  basis set that is energy independent;
(ii) normalization of the scattering path operator of EMTO is performed  
in a particular screened representation \cite{an.sa.00,vi.sk.00,vito.01}, 
a more involved procedure in 
comparison with the straight normalization of the NAO  basis set.

Moreover, it is important to keep in mind that the closure relations that are 
typically used  to build the matrix transformations are valid on
the Hilbert space spanned by the eigenvectors of the Hamiltonian. These
relations hold exactly for the numerical results of each code separately;
nevertheless, the numerical results produced by different codes are not 
identical in the mathematical sense. Small differences between the observables
computed with different codes may exist due to various factors --
from unwanted numerical roundoff errors to incompleteness of the basis sets.
This indicates that  the closure relation for the Hilbert space of the
EMTO  will not be exact (in the mathematical since) when used in the
Hilbert space spanned by the eigenvectors of {\sc Siesta}. Consequently,
we expect that the transformation matrix will fulfill the usual requirements
(i.e., unitarity) only within numerical inaccuracies.

In view of these formal difficulties in obtaining a basis transformation 
to match a multiple scattering method with a Hamiltonian based scheme 
we propose an approach using the fact that expectation values 
should be independent of the specific representation. We apply this 
fundamental concept of quantum mechanics to the orbital occupation matrix 
(density matrix, $n_{m_i, m_j}$) for the $d$-manifold 
(i.e., $m_i, m_j = 1, ...,5 $), and we look for a formally similar 
matrix transformation $W$: 
\begin{equation}
\begin{pmatrix}
W_{m_j,m_i}^{1\zeta}  &  
W_{m_j,m_i}^{2\zeta} 
\end{pmatrix} 
n^{\rm EMTO}_{m_im_j}
\begin{pmatrix}
W_{m_i,m_j}^{1\zeta} \\
W_{m_i,m_j}^{2\zeta} 
\end{pmatrix} = n^{\rm SIESTA}_{m_{i},m_{j}}
\label{transdef}
\end{equation}
Equation (\ref{transdef}) provides us with a system of equations to determine 
the matrix elements of $W$ numerically.
Indeed, by inspecting the diagonal elements of the occupation matrix for 
the $xy$, $yz$, $z^2-r^2$, $xz$ and $x^2-y^2$ orbitals respectively, we 
obtained the data shown in Table \ref{tab3}. One should note that in the 
physical-EMTO basis the occupation matrix is diagonal.  The corresponding
{\sc Siesta} occupation matrix has non-diagonal elements that are about two 
orders of magnitude smaller than the diagonal ones.  
\begin{table}[h]
\begin{tabular}{cc|c|c|c|c|c|l}
\cline{3-7}
%& & \multicolumn{4}{ c| }{Primes} \\ \cline{3-6}
& & $m_i=xy$ & $yz$ & $z^2-r^2$ & $zx$ & $x^2-y^2$ \\ \cline{1-7}
\multicolumn{1}{ |c| }{\multirow{2}{*}{$ n_{m_{i},m_{i}}^{EMTO}   $} } &
\multicolumn{1}{ |c| }{$\uparrow$} & 0.629 & 0.526 & 0.658 & 0.526 & 0.629   \\ \cline{2-7}
\multicolumn{1}{ |c  }{}                        &
\multicolumn{1}{ |c| }{$\downarrow$} & 0.922 & 0.925 & 0.908 & 0.925 & 0.922  \\ \cline{1-7}
\multicolumn{1}{ |c  }{\multirow{2}{*}{  $n_{m_{i},m_{i}}^{Siesta}$    } } &
\multicolumn{1}{ |c| }{$\uparrow$} & 0.639 & 0.485 & 0.784 & 0.485 & 0.639 \\ \cline{2-7}
\multicolumn{1}{ |c  }{}                        &
\multicolumn{1}{ |c| }{$\downarrow$} & 0.935 & 0.952 & 0.924 & 0.952 & 0.935 \\ \cline{1-7}
\end{tabular}
\caption{Occupation matrix of Co-d orbitals in the Co/Cu heterostructure}
\label{tab3}
\end{table}
While the symmetry and the qualitative trends in the occupations are the same, 
the exact numerical values are not. In other words, Eq. (\ref{transdef}) 
is an approximative (numerical) relation, however, the resulting 
$W$ matrix reflects the symmetry of $n_{m_{i},m_{i}}$. 
The non-zero elements on the diagonal are close 
to $1$ for the first-$\zeta$ block and take small imaginary values for 
the second-$\zeta$ one. The equations Eq.~(\ref{Wup}) and (\ref{Wdw}) 
provide explicit values.

Accordingly, given the matrix elements for the transformation matrix,  
the self-energy generated in the EMTO-basis set can be 
transferred to the double(multiple)-zeta {\sc Siesta}  basis set according to
\begin{equation} \label{usu}
\Sigma^{\rm SIESTA} = W \Sigma^{\rm EMTO} W^\dagger .
\end{equation}
In the following, we give simple examples to illustrate the self-energy 
transformation from the physical into a numerical double-zeta basis.

\subsection{Example: a two orbital model in the cubic symmetry}
\label{example2}
We consider a simplified case of a diagonal self-energy corresponding to 
a two orbital model in the cubic symmetry. The self-energy and the 
``occupation'' matrix can be written as:
\begin{equation}
\Sigma^{EMTO}(z) = 
\begin{pmatrix}
\Sigma_1 & 0  \\
0  & \Sigma_2 
\end{pmatrix} ; \ 
n^{EMTO} = \begin{pmatrix}
{n}_{1} & 0  \\
0  & {n}_{2} 
\end{pmatrix} .
\end{equation}
In such a case all inner products of the type 
$\langle \Psi_{m_i}(\epsilon, {\bf r})|\phi^{\zeta_1}_{m_j}(\bf r)\rangle$ 
are zero unless $m_i=m_j$, such that the transformation matrix has a generic 
form: 
\begin{equation}
W_{m_i,m_i}^{1\zeta,2\zeta} (\epsilon) = 
\begin{pmatrix}
W^{1\zeta}_{11} & 0  \\
0  & W^{1\zeta}_{22} \\
%\hline
W^{2\zeta}_{11}  & 0  \\
0  & W^{2\zeta}_{22}  \\
\end{pmatrix}.
\end{equation}
Re-writing Eq.~(\ref{transdef}):
\begin{equation}
\begin{pmatrix}
W^{1\zeta}_{11} & 0  \\
0  & W^{1\zeta}_{22} \\
W^{2\zeta}_{11}  & 0  \\
0  & W^{2\zeta}_{22}  \\
\end{pmatrix} 
\cdot 
 \begin{pmatrix}
{n}_{1} & 0  \\
0  & {n}_{2} 
\end{pmatrix}
\cdot 
\begin{pmatrix}
W^{1\zeta}_{11} & 0 & W^{2\zeta}_{11}  & 0  \\
0  & W^{1\zeta}_{22}& 0  & W^{2\zeta}_{22} 
\end{pmatrix} 
= 
\begin{pmatrix}
{\tilde n}_{1}^{1\zeta} & 0 & 0 & 0  \\
 0 &{\tilde n}_{2}^{1\zeta} & 0 & 0  \\
 0 & 0 &{\tilde n}_{1}^{2\zeta} & 0  \\
 0 &0 & 0 &{\tilde n}_{2}^{2\zeta}   \\
\end{pmatrix}
\end{equation}  
one finds  
\begin{equation}
W_{m_i,m_i}^{1\zeta,2\zeta} (\epsilon) =  
\begin{pmatrix}
\frac{\sqrt{ {\tilde n}_{1}^{1\zeta}  }}{\sqrt{n_{1}}}  & 0  \\
0  & \frac{\sqrt{ {\tilde n}_{2}^{1\zeta} }}{\sqrt{n_{2}}} \\
%\hline
\frac{\sqrt{ {\tilde n}_{1}^{2\zeta} }}{\sqrt{n_{1}}} & 0  \\
0  & \frac{\sqrt{ {\tilde n}_{2}^{2\zeta} }}{\sqrt{n_{2}}} \\
\end{pmatrix}, \ 
\end{equation}
and accordingly the self-energy in the NAO basis-set has the form:
\begin{equation}
\Sigma = \begin{pmatrix}
\Sigma_1 \frac{{ {\tilde n}_{1}^{1\zeta}}}{{n_{1}}} & 0 & \Sigma_1  \frac{ \sqrt{ {\tilde n}_{1}^{1\zeta}  {\tilde n}_{1}^{2\zeta}}}{n_{1}} &  0  \\
0  & \Sigma_2 \frac{ {\tilde n}_{2}^{1\zeta}}{n_{2}} & 0 & \Sigma_2 \frac{ \sqrt{ {\tilde n}_{2}^{1\zeta}  {\tilde n}_{2}^{2\zeta}}}{n_{2}}  \\
\Sigma_1  \frac{ \sqrt{ {\tilde n}_{1}^{1\zeta}  {\tilde n}_{1}^{2\zeta}}}{n_{1}} &  0 & \Sigma_1  \frac{{ {\tilde n}_{1}^{2\zeta}}}{{n_{1}}} &  0 \\
0 & \Sigma_2 \frac{ \sqrt{ {\tilde n}_{2}^{1\zeta}  {\tilde n}_{2}^{2\zeta}}}{n_{2}} & 0 & \Sigma_2  \frac{ {\tilde n}_{2}^{2\zeta}}{n_{2}} 
\end{pmatrix} = 
\begin{pmatrix}
\fbox{$\Sigma^{1\zeta,1\zeta}$} & \fbox{$\Sigma^{1\zeta,2\zeta}$} \\
\fbox{$\Sigma^{2\zeta,1\zeta}$} & \fbox{$\Sigma^{2\zeta,2\zeta}$} 
\end{pmatrix}
\end{equation} 
in which every $\Sigma^{i\zeta,j\zeta}$ is a $2 \times 2 $ 
block-diagonal matrix. 
There are a couple of conclusions to be drawn from the above
simplified example: (i) as a consequence of the reduced weight in the  
second $\zeta$ occupation (at least 10 times smaller that in the single 
$\zeta$), the magnitude of the matrix elements of the self-energy in the 
NAO  basis set follow the relation
$ \Sigma^{1\zeta,1\zeta} > \Sigma^{1\zeta,2\zeta} > \Sigma^{2\zeta,2\zeta}$;
(ii) the existence of only-diagonal orbital occupations does not imply the 
existence of an unitary transformation, except for the case of numerical 
identical values for the occupation matrices in both basis. These observations 
do not change when the symmetry is lower than cubic and non-zero matrix 
elements on the non-diagonal of the occupation matrix appear.

\section{Transformation matrix for the Co-d manifold in the Cu-Co-Cu 
heterostructure}
This section provides the spin-resolved matrix elements, 
$W_{\sigma=\uparrow, \downarrow}$, used in the calculation.

%%%%%%%%%%%%%%%%%%%%%%%%%%%
\begin{equation}
W_{\uparrow}= \left(
\begin{array}{ccccc}
 1.024 & 0 & 0 & 0 & 0 \\
 0 & 0.987 & 0 & 0 & 0 \\
 0 & 0 & 1.096 & 0 & 0 \\
 0 & 0 & 0 & 0.985 & 0 \\
 0 & 0 & 0 & 0 & 1.024 \\
 0.178 i & 0 & 0 & 0 & 0 \\
 0 & 0.218 i & 0 & 0 & 0 \\
 0 & 0 & 0 & 0 & 0 \\
 0 & 0 & 0 & 0.218 i & 0 \\
 0 & 0 & 0 & 0 & 0.178 i \\
\end{array}
\right ). \label{Wup}
\end{equation}
The orthogonality can be checked using the following relations:
\begin{equation}
W_{\uparrow}^\dagger W_{\uparrow} = 
\left (
\begin{array}{ccccc}
 1.016 & 0 & 0 & 0 & 0 \\
 0 & 0.922 & 0 & 0 & 0 \\
 0 & 0 & 1.202 & 0 & 0 \\
 0 & 0 & 0 & 0.922 & 0 \\
 0 & 0 & 0 & 0 & 1.016 \\
\end{array}
\right )
\end{equation}
and 
\begin{equation}
W_{\uparrow} W_{\uparrow}^{^\dagger} = 
\left(
\begin{array}{cccccccccc}
 1.048 & 0 & 0 & 0 & 0 & 0.182 i & 0 & 0 & 0 & 0 \\
 0 & 0.970 & 0 & 0 & 0 & 0 & 0.215 i & 0 & 0 & 0 \\
 0 & 0 & 1.202 & 0 & 0 & 0 & 0 & 0 & 0 & 0 \\
 0 & 0 & 0 & 0.970 & 0 & 0 & 0 & 0 & 0.215 i & 0 \\
 0 & 0 & 0 & 0 & 1.047 & 0 & 0 & 0 & 0 & 0.182 i \\
 0.182 i & 0 & 0 & 0 & 0 & -0.032 & 0 & 0 & 0 & 0 \\
 0 & 0.215 i & 0 & 0 & 0 & 0 & -0.048 & 0 & 0 & 0 \\
 0 & 0 & 0 & 0 & 0 & 0 & 0 & 0 & 0 & 0 \\
 0 & 0 & 0 & 0.215 i & 0 & 0 & 0 & 0 & -0.048 & 0 \\
 0 & 0 & 0 & 0 & 0.183 i & 0 & 0 & 0 & 0 & -0.032 \\
\end{array}
\right )
\end{equation}
The corresponding transformation matrix for spin-down component
reads:
\begin{equation}
W_{\downarrow} =\left(
\begin{array}{ccccc}
 0.960 & 0 & 0 & -0.007 & 0 \\
 0 & 0.9572 & 0 & 0 & -0.007 \\
 0 & 0 & 0.967 & 0 & 0 \\
 -0.007 & 0 & 0 & 0.957 & 0 \\
 0 & -0.007 & 0 & 0 & 0.960 \\
 0.304 & 0 & 0 & -0.002 & 0 \\
 0 & 0.337 & 0 & 0 & -0.002 \\
 0 & 0 & 0.287& 0 & 0 \\
 -0.002 & 0 & 0 & 0.337 & 0 \\
 0 & -0.002 & 0 & 0 & 0.304 \\
\end{array}
\right ) \label{Wdw}
\end{equation}

\begin{equation}
W_{\downarrow}^\dagger W_{\downarrow} = \left(
\begin{array}{ccccc}
 1.014 & 0& 0& -0.015 & 0\\
 0& 1.029 & 0& 0& -0.015 \\
 0& 0& 1.017 & 0& 0\\
 -0.015 & 0& 0& 1.029 & 0\\
 0& -0.015 & 0& 0& 1.014 \\
\end{array}
\right )
\end{equation}

\begin{equation}
W_{\downarrow} W_{\downarrow}^\dagger =
\left(
\begin{array}{cccccccccc}
 0.922 & 0& 0& -0.014 & 0& 0.292 & 0& 0& 0& 0\\
 0& 0.916 & 0& 0& -0.014 & 0& 0.322 & 0& 0& 0\\
 0& 0& 0.935 & 0& 0& 0& 0& 0.278 & 0& 0\\
 -0.014 & 0& 0& 0.916 & 0& 0& 0& 0& 0.322 & 0\\
 0& -0.014 & 0& 0& 0.922 & 0& 0& 0& 0& 0.292 \\
 0.297 & 0& 0& 0& 0& 0.092 & 0& 0& 0& 0\\
 0& 0.325 & 0& 0& 0& 0& 0.114 & 0& 0& 0\\
 0& 0& 0.278 & 0& 0& 0& 0& 0.083 & 0& 0\\
 0& 0& 0& 0.325 & 0& 0& 0& 0& 0.114 & 0\\
 0& 0& 0& 0& 0.292 & 0& 0& 0& 0& 0.092 \\
\end{array}
\right )
\end{equation}

\section{Assessment of accuracy}

\begin{figure}[h]
\includegraphics[width=0.5\columnwidth]{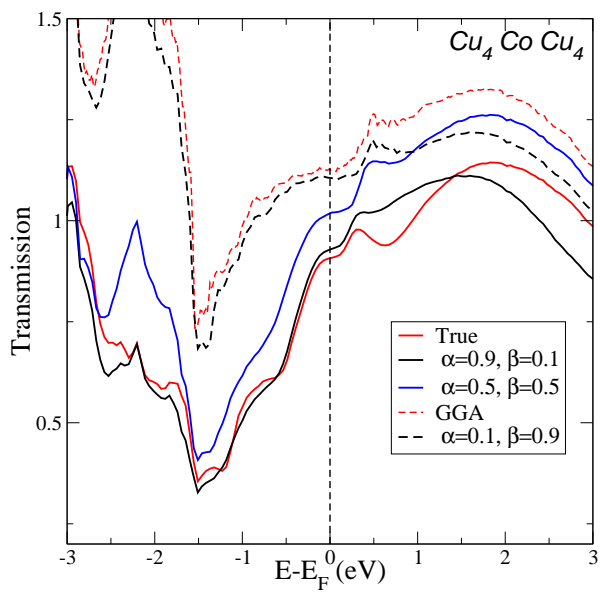}
\caption{(color online) Comparison between the transmission functions 
obtained for different model-forms  of $W^{1\zeta, 2\zeta}$. } 
\label{fig7}
\end{figure}

In order to test the effect of the numerical inaccuracies occurring in 
the transformation matrix on the final results, we compute the total 
transmission by using a simplified model for the transformation matrix:
\begin{equation}
W^{1\zeta, 2\zeta}_{Model} =
\begin{pmatrix}
\fbox{$ W^{1\zeta}$}  \\
\fbox{$ W^{2\zeta}$}
\end{pmatrix},
\end{equation}
where $W^{1\zeta} = \alpha I $ and $W^{2\zeta} = \beta I$, where 
$I$ is the unity matrix. The results for $\alpha = 0.9 $, 
$\beta= 0.1$; $\alpha = 0.5 $, $\beta= 0.5$; and  $\alpha = 0.1 $, 
$\beta= 0.9$, respectively are given  in Fig.~\ref{fig7}. 
It can be clearly seen that even for such a crude approximation,
the results for the first model (i.e., with a significant weight of 
the self-energy on the first zeta orbital) differ with only a few 
percent over large energy domains. For the 
occupied states, large values for $\Sigma^{1\zeta} $ provide already
a good approximation for the transformation.

\bibliography{references_database}
\end{document}